\documentclass[12pt]{article}
 \usepackage{amssymb}
 \usepackage{amsbsy}
 \usepackage{amscd}
 \usepackage{amsmath}
 \usepackage{amsthm}
\newcommand{\RR}{\mathbb R}
\newcommand{\CC}{\mathbb C}
\newcommand{\EE}{\mathbb E}
\newcommand{\VV}{\mathbb V}

\title{Review and Prospect of Algebraic Research in 
Equivalent Framework between
Statistical Mechanics and Machine Learning Theory}
\author{Sumio Watanabe\\
RIKEN Center for Advanced Intelligence Project\\
E-mail:sumio.watanabe@riken.jp
}
\date{}

\begin{document}

\maketitle

\begin{abstract}
Mathematical equivalence between statistical mechanics and machine learning theory has been known since the 20th century, and research based on this equivalence has provided novel methodologies in both theoretical physics and statistical learning theory. It is well known that algebraic approaches in statistical mechanics such as operator algebra enable us to analyze phase transition phenomena mathematically. In this paper, we review and prospect algebraic research in machine learning theory for theoretical physicists who are interested in artificial intelligence.

If a learning machine has a hierarchical structure or latent variables, 
then the random Hamiltonian cannot be expressed by any quadratic perturbation because it has singularities. To study an equilibrium state defined by such a singular random Hamiltonian, algebraic approaches are necessary to derive the asymptotic form of the free energy and the generalization error. 

We also introduce the most recent advance: the theoretical foundation for the alignment of artificial intelligence is now being constructed based on algebraic learning theory.

This paper is devoted to the memory of Professor Huzihiro
Araki who is a pioneering founder of algebraic research  in both 
statistical mechanics and quantum field theory. 
\end{abstract}

\section{Introduction}

Statistical mechanics was founded in
the 19th century and have evolved throughout the 20th and 21th centuries. 
Information theory was established in the 20th century \cite{Shannon1948}, and formal similarity between statistical mechanics and 
information theory has been sometimes pointed out 
\cite{Jaynes1957, Brillouin1956,Landauer1961,Bennett1973}. 
Even if nature and principle of two fields are different, 
such formal similarity can be treated by common mathematical
foundation.

In the second half of the 20th century, 
equivalent structure was found between statistical mechanics of random
interactions and learning and memory in artificial neural networks
\cite{Edwards1975,Sherrington1975,Parisi1980,
Hopfield1982,Mezard1987,Levin1990,Talagrand2000}.
The extensive researches in these field  
became one of the foundations of today's artificial intelligence. 

Among mathematical physics in
 statistical mechanics and quantum field theory, 
algebraic approach was indicated in the 1960s by pioneer work 
by Huzihiro Araki \cite{Araki1964,Araki1968,Araki1974}.
In fact, types of von Neumann algebra made by a spin system was
classified, quantum field theory can be treated by axiomatic way, 
and mathematical 
equivalence of equilibrium state and variational principle was 
clarified.

In this paper, for theoretical physicists who are interested in
artificial intelligence, 
we review and prospect algebraic researches in
machine learning theory and its relation to artificial 
intelligence alignment. We expect that machine learning theory
would be extended from the viewpoint of statistical physics. 

This paper consists of four sections. In the second section, 
we explain the well-known formal equivalence of statistical physics 
and machine learning theory. In subsection 2.1, 
structure of statistical physics is described 
in a manner familiar for physicists, however, this subsection 
is prepared for comparison with
the following subsections. In subsection 2.2, structure of 
machine learning theory is introduced. Physicists 
understand that machine learning theory is equivalent to
statistical mechanics with random Hamiltonian. 
In subsection 2.3,  we explain how concepts in statistical mechanics
correspond to ones in machine learning theory.
In the third section, we review and prospect 
algebraic approach in machine learning theory.
If a learning machine contains hierarchical structure or hidden variables,
then the random Hamiltonian has singularities, resulting that 
algebraic approach is necessary. 
In subsection 3.1, the main results in algebraic approach of 
machine learning theory are summarized. It is shown that
asymptotic form of the free energy and generalization error 
are represented by two birational invariants, the real log canonical
threshold and singular fluctuation. 
In subsection 3.2, mathematical properties of real log canonical threshold
are clarified and its application to estimation of the free energy is explained. 
In subsection 3.3, mathematical properties of singular fluctuation  
are studied and its application to estimation of the generalization error
is examined. 
In subsection 3.4, the most recent researches about artificial intelligence
alignment is prospected. We expect that mathematical researches 
about statistical physics and machine learning theory would be 
a foundation of the future artificial intelligence development.

\section{Equivalent Framework of Statistical Mechanics and 
Statistical Learning Theory}

In this section we summarize the well-known common framework of 
statistical mechanics and statistical learning theory
\cite{Edwards1975,Sherrington1975,Parisi1980,
Hopfield1982,Mezard1987,Levin1990,Talagrand2000}. 
Even if statistical mechanics and statistical learning theory 
are based on different nature and principle, 
their formal structure can be studied based on a  
common mathematical framework, hence, from the mathematical
point of view, there is no need to distinguish them. 
The correspondence between statistical mechanics and
machine learning theory is summarized in Table \ref{table:111}.

\begin{table}
\begin{center}
\begin{tabular}{|c|c|c|c|}
\hline
 \multicolumn{2}{|c|}{ Statistical Mechanics} &
  \multicolumn{2}{|c|}{ Machine Learning Theory}
 \\
\hline
$J$ & random interaction & $X^n$ 
& random data\\
\hline
$w$ & observable  & $w$ & parameter of model\\
\hline
$H(w,J)$ & random Hamiltonian & $H(w,X^n)$ &minus log likelihood
\\
 \hline
 $\varphi(w)$& weight of $w$ & $\varphi(w)$ &prior distribution of $w$
 \\
 \hline
 $\rho(w|J)$ 
 & equilibrium state & $\rho(w|X^n)$ & posterior distribution 
  \\
 \hline
 $Z(\beta,J)$& partition function & $Z(\beta,X^n)$ & marginal likelihood 
 \\
 \hline
 $F(\beta,J)$& free energy& $F(\beta,X^n)$ & free energy  \\
 \hline
\end{tabular}
\end{center}
\caption{Statistical Mechanics and Learning Theory\\
This table shows the common mathematical framework of statistical mechanics and
machine learning theory. Both $H(w,J)$ and $H(w,X^n)$ are made as candidate models,
whose validations are performed by 
 comparing theoretical and observed free energies.}
\label{table:111}
\end{table}

\subsection{Statistical Mechanics of Random Hamiltonian}

First, we explain the formal framework of statistical mechanics,
which is well-known for statistical physicists; 
this subsection is necessary for comparing with the following 
subsections. 

Let $J$ be a random variable defined on a finite
dimensional Euclidean space and $w\in\RR^d$ be an observable. 
In order to examine a physical system with random interaction, 
a function $H(w,J)$ of $(w,J)$ is set as a candidate model, which
is referred to as {\it a random Hamiltonian}. 
For a given random variable $J$, 
the equilibrium state with an inverse temperature $\beta>0$
is defined by a probability density function of $w$, 
\begin{align}\label{eq:boltzmann}
\rho(w|J)=\frac{1}{Z(\beta,J)}\exp(-\beta H(w,J))\varphi(w),
\end{align}
where $\varphi(w)$ is a nonnegative function of $w$ and 
$Z(\beta,J)$ is the normalizing constant, 
\[
Z(\beta,J)=\int\exp(-\beta H(w,J))\varphi(w)dw. 
\]
If the principle of equal weights is adopted, then $\varphi(w)\equiv 1$. 
The normalizing constant $Z(\beta,J)$ is called a {\it partition function}. 
The expectation and variance with respect to the probability 
distribution $\rho(w|J)$ 
are denoted by 
$\EE_w^\beta[\;\;]$ and $\VV_w^\beta[\;\;]$, respectively. 
The {\it free energy} is defined by
\[
F(\beta,J)=-\frac{1}{\beta}\log Z(\beta,J).
\]
The average free energy over $J$ is denoted by $\EE_J[F(\beta,J)]$. 
It is well known in statistical mechanics that, 
if the average free energy is explicitly calculated as a function of 
$\beta$, 
then several important theoretical values in physics can be derived from the 
average free energy.
Validation of modeling a random Hamiltonian is performed by
comparing the theoretical results obtained from the average 
free energy with the experimental observation. 

\vskip3mm\noindent{\bf Example in Statistical Mechanics}. 
Let $J=\{J_{ij}\}$ and $w=\{w_i\}$, where $w_i=\pm 1$ or 
some continuous numbers. For 
a probability distribution of $J$, 
a normal distribution or another one is sometimes employed.  
A physical model of a spin glass system \cite{Edwards1975,Sherrington1975} is defined by 
a random Hamiltonian, 
\begin{align}
H(w,J)&=-\sum_{\{i,j\}} J_{ij} w_i w_j-\sum_{i}h_i w_i
\label{eq:spin}
\end{align}
where $h=\{h_i\}$ is 
the external magnetic field. If the principle of equal wights is 
adopted, $\varphi(w)\equiv 1$. Then the conditional
distribution $\rho(w|J)$ represents Boltzmann distribution of
the equilibrium state with inverse temperature $\beta>0$. 
If the average free energy is calculated, 
the average magnetization 
is obtained from the free energy, 
\[
M(\beta)=\EE_{J}\EE_w^\beta\left[
\frac{1}{n}\sum_i w_i  
 \right]
= \frac{1}{n}\sum_i\frac{\partial}{\partial h_i}\EE_{J}[F(\beta,J)].
\]
The appropriateness of a model given by eq.(\ref{eq:spin}) can 
be validated by comparing the average magnetization $\chi(\beta)$ 
with experimental results. 

\subsection{Statistical Learning Theory}

Second, we explain machine learning theory. 
Statistical physicists will be able to see that 
machine learning theory is formally equivalent to 
statistical mechanics. 

Let $d$, $n$, $N$ are positive finite integers. 
In stead of a random interaction $J$, we use a notation for
a random sample $X^n$. 
Let 
\[
X^n\equiv\{X_i\;;\;i=1,2,...,n\}
\]
be a set of $\RR^N$ valued random variables which are 
independently subject to an unknown 
 probability density function $q(x)$.
A learning machine is defined by a conditional 
probability density function $p(x|w)$ and a 
probability density function $\varphi(w)$ where 
 $x\in\RR^N, w\in \RR^d$. 
In statistical learning theory, 
$q(x)$ is an unknown data-generating distribution,
and a pair of $p(x|w)$ and $\varphi(w)$ is a candidate model which is 
designed by a user. There is no guarantee that 
a model will be appropriate for an unknown data-generating
process, hence we need mathematical theory which holds
 for an arbitrary triple $(q(x),p(x|w),\varphi(w))$. 

A random Hamiltonian 
\begin{align}
H(w,X^n)=-\sum_{i=1}^n \log p(X_i|w),
\label{eq:Hwx}
\end{align}
is called the {\it minus log likelihood} in learning theory. 
The probability density function 
\begin{align}
\rho(w|X^n)& =\frac{1}{Z(\beta,X^n)}\varphi(w)\prod_{i=1}^n p(X_i|w)^\beta
\nonumber
\\
&=\frac{1}{Z(\beta,X^n)}\varphi(w)\exp(-\beta H(w,X^n))
\end{align}
is called a {\it posterior distribution}, where
\[
Z(\beta,X^n)=\int \varphi(w)\prod_{i=1}^n \exp(-\beta H(w,X^n)) dw.
\]
This definition is equal to eq.(\ref{eq:boltzmann}). 
The expectation and variance using $\rho(w|X^n)$ are
also denoted by $\EE_w^\beta[\;\;]$ and 
$\VV_w^\beta[\;\;]$, respectively. 
The inference methods $\beta=1$ and $\beta=\infty$ are called
{\it Bayesian} and {\it maximum likelihood} ones, respectively.
The  free energy of a learning machine is 
given by 
\[
F(\beta,X^n)=-\frac{1}{\beta}\log Z(\beta,X^n). 
\]
Then random variables $Z(1,X^n)$ and $F(1,X^n)$ are
referred to as the {\it marginal likelihood} and
the {\it minus log marginal likelihood}, both of which
are important values in statistics. Also
$F(1,X^n)$ is called {\it stochastic complexity} or
{\it Bayesian code length} in information theory. 
Note that $Z(1,X^n)$ is a probability density function of $X^n$ 
defined by a model and a prior, 
because 
\[
\int Z(1,x^n) dx^n=1
\]
and the average free energy is equal to the sum of the 
entropy and the relative entropy. 
\[
\EE_{X^n}[F(1,X^n)]=nS+
\int q(x^n) 
\log\left(\frac{q(x^n)}{
Z(1,X^n)}
\right)dx^n
\]
where $S$ is the entropy of the unknown data-generating distriution 
\[
S=-\int q(x)\log q(x)dx
\]
and $q(x^n)=\prod_{i=1}^nq(x_i)$. This equation shows that
the smaller free energy is equivalent to the smaller relative 
entropy between $q(x^n)$ and $Z(1,X^n)$. 
The {\it posterior predictive distribution} is defined by the average of 
$p(x|w)$ using the posterior distribution 
\[
p(x|X^n)=\EE_{w}^\beta[ p(x|w) ],
\]
which shows the conditional probability function of $x$ for
a given $X^n$ using $p(x|w)$ and $\varphi(w)$. 
Then the {\it generalization error} is defined by the relative
entropy of $q(x)$ and $p(x|X^n)$,
\[
G_n(\beta)=\int q(x)\log\frac{q(x)}{p(x|X^n)}dx,
\]
which measures a difference between the unknown data-generating 
distribution $q(x)$ and the posterior predictive distribution  $p(x|X^n)$. 
The smaller generalization error shows the smaller relative entropy of
$q(x)$ and $p(x|X^n)$, which shows the a model and a prior are 
more appropriate for the unknown data-generating distribution
 according to prediction accuracy. 
Hence, one of the most important purposes in statistical
learning theory is to clarify the behavior of the generalization 
error $G_n(\beta)$ as a function of $n$. 
However, the integration over $q(x)$ in the definition 
$G_n(\beta)$ cannot be performed in real applications 
because $q(x)$ is unknown. On the other hand, 
the {\it training error}
\[
T_n(\beta)=\frac{1}{n}\sum_{i=1}^n \log \frac{q(X_i)}{p(X_i|X^n)}
\]
is used to estimate the generalization error. 
One of the purposes of statistical learning theory is to derive 
the difference 
between the generalization and training errors
\begin{align}\nonumber
&\EE_{X^n}[G_n(\beta)-T_n(\beta)]
\\
&=\EE_{X^n}\left[
\int q(x)\log p(x|X^n)dx-\frac{1}{n}
\sum_{i=1}^n\log p(X_i|X^n)
\right],
\end{align}
which is not equal to zero in general because 
$X_i$ and $X^n$ are not independent.
 If $\beta=1$, then by the definition of the predictive distribution, 
\[
p(X_{n+1}|X^n)=\frac{Z(1,X^{n+1})}{Z_n(1,X^n)},
\]
resulting that 
\begin{align}
\EE_{X^n}[G_n(1)]&=
\EE_{X^{n+1}}[F(1,X^{n+1})-F(1,X^n)]-S.
\label{eq:diff_free}
\end{align}
The relation eq.(\ref{eq:diff_free}) shows that the 
generalization error for $\beta=1$ 
is derived from the free energy if it is
given by an explicit function of $n$. 
In machine learning, for the validation of a candidate model and 
prior $p(x|w)$ and $\varphi(w)$, 
$F(1,X^n)$ and $G_n(1)$ are examined. 
Although $F(1,X^n)$ and $G_n(1)$ have a mathematical
relation eq.(\ref{eq:diff_free}), the pair of model and prior that 
minimizes $F(1,X^n)$ is different from one that minimizes
$G_n(1)$, in general. 

\vskip3mm\noindent{\bf Example in Machine Learning}.
Let $x=(x_1,x_2)\in\RR^M\times \RR^N$ 
and assume that $\{(X_{1i},X_{2i});i=1,2,...n\}$ is a set of 
independent random variables whose probability density function
is $q(x_1)q(x_2|x_1)$. A learning machine is sometimes employed 
\begin{align}\label{eq:f(x,w)}
p(x_1,x_2|w1,\sigma)=\frac{q(x_1)}{(2\pi\sigma^2)^{N/2}}
\exp\left(
-\frac{1}{2\sigma^2}(x_2-f(x_1,w_1))^2\right),
\end{align}
where $f(x,w_1)$ is a function from $\RR^M$ to $\RR^N$ which has 
a parameter $w_1$. 
In deep learning, a function $f(x,w_1)$ is defined by using a layered
neural network and the parameter is $w=(w_1,\sigma)$. 
Remark that, in this model, an unknown conditional
probability distribution $q(x_2|x_1)$ is estimated 
whereas $q(x_1)$ is not estimated.

\subsection{Correspondence between Statistical Mechanics
 and Machine Learning}

In this subsection, we summarize the correspondence between
statistical mechanics and machine learning, which is displayed in 
Table \ref{table:111}. There are much more relations than this table 
between 
statistical mechanics and machine 
learning theory. The mean field approximation in the statistical mechanics 
is equivalent to 
{\it variational Bayesian inference} in machine learning \cite{WatanabeK2006,Nakajima2019},
The Langevin equation in statistical mechanics is just equal to 
the steepest descent with random noise in neural network learning, 
and the probability distribution of the random free energy gives the 
foundation of design of the 
most powerful test in statistical hypothesis testing \cite{Kariya2022}.

In statistical mechanics,
the random Hamiltonian is devised as a candidate model 
for studying a phenomenon which has random interaction
 such as spin glass systems. 
Sometimes an assumption is set that 
 $J$ is subject to normal distribution, 
validation of which is performed by
comparing theoretical values with experimental observation. 
In statistical mechanics, even if a candidate model
cannot describe microscopic interaction in detail,
 renormalizing enables us to understand 
 the characteristics of macroscopic phenomena. 
This property is called universality in theoretical physics. 

In Bayesian statistics \cite{GelmanBDA,Gelman2013,McElreath2020}, 
a pair of $p(x|w)$ and $\varphi(w)$ 
is a candidate model for estimating an unknown uncertainty or
an unknown information source $q(x)$. 
In statistics, it is premised that all models are wrong 
\cite{Box1976,Binmore2017} and the more 
appropriate pair of $p(x|w)$ and $\varphi(w)$ depends on the 
sample size $n$. 
If a pair, $p(x|w)$ and $\varphi(w)$, 
is designed, then $X^n$ is a set of 
exchangeable random variables. 
For exchangeable random variables,
by de Finetti-Hewitt-Savage theorem 
\cite{Hewitt1955}, there exists a functional probability distribution
$Q$ such that
\begin{align}
q(x)&\sim Q,
\\
X^n&\sim\prod_{i=1}^n q(x_i).
\end{align}
Note that $(1/n)\sum_{i=1}^n X_i$ converges to 
$\int x q(x)dx$ which depends on $q(x)$, where $q(x)$ is
a function-valued random variable. Therefore, 
if a user of Bayesian statistics or a scientist 
designs a pair, $p(x|w)$ and $\varphi(w)$, as a candidate for an 
unknown uncertainty, it is also assumed that 
there exists unknown functional probability distribution $Q$ 
to which $q(x)$ is subject
\cite{Watanabe2023}. 
Validation of the candidate pair for unknown uncertainty is examined by comparing
the theoretical values and experimental observation. 

In statistical mechanics, thermodynamical limit sometimes plays
important roles, by which a macroscopic phenomenon is 
derived from a microscopic stochastic dynamics. For example, 
the dimension of $w=\{w_i\}$ is made to be infinity 
in spin glass theory. Also in machine learning theory, 
thermodynamical limit can be taken if the leaning machine 
has homogeneous structure. In fact, if the function $f(x_1,w_1)$ 
in eq.(\ref{eq:f(x,w)}) is given by the inner product of $x_1$ and $w_1$,
\[
f(x_1,w_1)=(x_1)\cdot(w_1)
\]
then it is possible to study thermodynamical limit.
 However, for non-homogeneous physical phenomena such as Bose-Einstein
 condensation, it seems to be impossible to study thermodynamical limit. 
 In this paper we study nonhomogeneous learning machines such
 as hierarchical and degenerate neural networks which contain 
 complex singularities, resulting that there may not exist the thermodynamical
 limit. It would be an important problem for the future study how to examine
 the large scale limit of nonhomogeneous learning machines without using 
 thermodyamical limit. In statistical learning theory, 
almost all learning machines are singular \cite{Watanabe2007}
and that's good \cite{Wei2022}, which seems to be one of the most 
 essential  properties of brain-like inference systems.

\section{Algebraic Researches in Statistical Learning Theory}

In this chapter we review and prospect 
algebraic researches in statistical learning theory.

\subsection{Mathematical Results}

\label{subsection:mainresult}

In this subsection, we explain the main results of singular   
learning theory. 

Assume that a prior distribution, $\varphi(w)$, is a $C_0^\infty$ class function 
of $\RR^d$ and that 
$W\subset\RR^d$ is a compact subset whose open kernel
contains the support of $\varphi(w)$, $\mbox{supp}\;\varphi$.
Let us define a function of $W$ 
\[
L(w)=-\int q(x) \log p(x|w) dx.
\]
Then the expectation value of the random Hamiltonian 
eq.(\ref{eq:Hwx}) is 
\[
\EE_{X^n}[H(w,X^n)]=n L(w). 
\]
Note that $L(w)\geq S$, where $S$ is the entropy of $q(x)$. 
We assume that $L(w)$ is a real analytic function on the open
kernel of $W$ 
and that there exists $w_0\in\mbox{supp}\;\varphi$ 
which minimizes $L(w)$ $(w\in W)$. 
The set of all parameters which minimize $L(w)$ in $W$ is denoted by
\[
W_0\equiv \{w\in W\;;\; L(w)=L(w_0)\}.
\]
In other words, $W_0$ consists of all zero points of the 
real analytic function $L(w)-L(w_0)$, which is called an {\it analytic set}. 
If $w_0$ satisfies 
$q(x)=p(x|w_0)$, then $q(x)$ is said to be {\it realizable} by $p(x|w)$. 
Note that, if $q(x)$ is realizable by $p(x|w)$, then $L(w_0)$ is
equal to the entropy $S$ of $q(x)$. 
If there exists a unique 
$w_0$ 
that minimizes $L(w)$ and if the Hessian matrix 
$\nabla^2 L(w_0)$ 
is positive definite, then $q(x)$ is said to be {\it regular} for $p(x|w)$; 
otherwise it is said to be {\it singular} for $p(x|w)$.

If $q(x)$ is regular for $p(x|w)$, 
 then the posterior distribution concentrates 
in the neighborhood of $w_0$ as $n\rightarrow\infty$, resulting that
the posterior distribution can be asymptotically approximated by a normal
distribution,
\[
\rho(w|X^n)\approx (1/C) \exp\left(
-\frac{\beta n}{2} (w-w^*)(\nabla^2 L(w_0))(w-w^*)\right),
\]
where $w^*$ is the unique parameter that minimizes $H(w,X^n)$ 
and $C$ is a normalizing constant. 
The conventional theory for  regular  cases is called
{\it regular learning theory}. 

Statistical learning theory 
that for singular case is called {\it singular  
learning theory}. From the mathematical point of view,
singular learning theory contains regular learning theory as a
very special example, hence the former is a generalization of
the latter. 
 If a learning machine contains latent variables or
hierarchical structure, then the regularity condition is not 
satisfied
\cite{Hartigan1985,Hagiwara1993,Fukumizu1996}, 
hence we need  singular learning theory both in statistics and 
machine learning. 
For example, in deep learning, the rank of 
the Hessian matrix in the neighborhood of the obtained 
parameter by training is far smaller than the dimension of $w$ in general. 

Let $w_0$ be an element of the set $W_0$. 
We define a function 
\[
K(w)=L(w)-L(w_0). 
\]
We need a method to analyze the set 
 \[
 W_\varepsilon\equiv \{w\in W\;;\; K(w)\leq \varepsilon\}
 \]
for a real analytic function $K(w)$. In general, the set
$W_0$ contains singularities, hence we need 
singularity theory to study $W_0$. 
Such a method was constructed based on 
singularity theory, algebraic geometry, and algebraic analysis 
 \cite{Atiyah1970,Kashiwara1976}.
The following theorem is the foundation 
of  singular learning theory. 
\vskip3mm\noindent
{\bf Hironaka Resolution Theorem} \cite{Hironaka1964}. 
For a given real analytic function 
$K(w)\geq 0$,
there exist both 
a compact subset ${\cal M}$ of a $d$-dimensional analytic manifold and 
a {\it proper} real analytic function from ${\cal M}$ to $W$ 
\[
g:{\cal M}\ni u \mapsto g(u)\in W
\]
such that, in each local coordinate of ${\cal M}$, 
$K(g(u))$ is normal crossing, in other words,
\begin{align}
K(g(u))&=u_1^{2k_1}u_2^{2k_2}\cdots u_d^{2k_d},
\label{eq:res01}
\\
\varphi(g(u))|g'(u)|&= b(u)|u_1^{h_1}u_2^{h_2}\cdots u_d^{h_d}|, 
\label{eq:res02}
\end{align}
where $k=(k_1,k_2,...,k_d)$ and $h=(h_1,h_2,...,h_d)$ are 
multi-indices of nonnegative integers, in which 
at least one $k_i$ is a positive integer. Here 
$b(u)>0$ is a positive analytic function, and $|g'(u)|$ is the
absolute value of the Jacobian determinant of $w=g(u)$. 
The correspondence between $W\setminus W_0$ and 
$g^{-1}(W\setminus W_0)$ is one-to-one.
Note that a function $w=g(u)$ is called {\it proper} 
if the inverse image of a compact set is also compact.
\vskip3mm\noindent
Concerning eqs. (\ref{eq:res01}) and (\ref{eq:res02}), 
we introduce 
the {\it real log canonical threshold} (RLCT) $\lambda$ and 
{\it multiplicity} $m$ by
\begin{align}
\lambda&=\min_{L.C.}\min_{1\leq j\leq d}\left(\frac{h_j+1}{2k_j}\right),
\\
m&=\max_{L.C.}\#\left\{j;\frac{h_j+1}{2k_j}=\lambda\right\},
\end{align}
where $\min_{L.C.}$ and $\max_{L.C.}$ denote 
 the minimum and maximum values over all local coordinates, respectively.
Here we define $(h_j+1)/(2k_j) =\infty$ for $k_j=0$,
and $\#$ is the number of elements of a set. 
Hence $0<\lambda<\infty$ and $1\leq m\leq d$. 
Since $W$ is compact and $w=g(u)$ is proper, the number of all 
local coordinates can be taken to be finite. 
The {property and application of RLCT is discussed in subsection
\ref{subsection:RLCT}.

Additionally we assume the relatively finite condition 
that, there exists $c_0>0$ such that,
for an arbitrary $w\in W$, 
\begin{align}\label{eq:finite}
\int q(x) f(x,w) dx\geq c_0 \int q(x) f(x,w)^2 dx,
\end{align}
where $f(x,w)=\log(p(x|w_0)/p(x|w))$. Since $W$ is a compact set,
both sides of eq.(\ref{eq:finite}) are trivially finite. 
By using $K(w)=\int q(x)f(x,w)dx$, 
eq.(\ref{eq:finite}) is equivalent to the condition that the variance of $f(x,w)$ is 
bounded by its average in a set $\{w;L(w)<\varepsilon\}$ for a sufficiently small
$\varepsilon>0$. In the case when this condition is not
satisfied, the variance of random Hamiltonian cannot be bounded by 
its average, resulting that the free energy has a different 
asymptotic behavior \cite{Nagayasu2022}.
By using resolution theorem, when the condition eq.(\ref{eq:finite}) is
satisfied, 
it is proved that the following limit is finite
\begin{align}\label{eq:nu(beta)}
\nu(\beta)=\frac{\beta}{2}\lim_{n\rightarrow\infty}
n\; \EE_{X^n}\left[ \VV_w^\beta[\log p(X_i|w)]
\right],
\end{align}
which is referred to as {\it  singular  fluctuation}. 
The property and application of  singular fluctuation
 is discussed in subsection
\ref{subsection:SF}.

The main results of 
 singular learning theory \cite{Watanabe2009,Watanabe2018} are 
 described as follows. 
\begin{align}
\EE_{X^n}[F_n(\beta)]&=nK(w_0)+nS+\frac{\lambda}{\beta} \log n 
-\frac{m-1}{\beta}\log\log n +O(1),
\\
\EE_{X^n}[G_n(\beta)]&=K(w_0)+\frac{1}{n}
\left(\frac{\lambda-\nu(\beta)}{\beta} + \nu(\beta)
\right)+o\left(\frac{1}{n}\right),
\\
\EE_{X^n}[T_n(\beta)]&=
K(w_0)+\frac{1}{n}
\left(\frac{\lambda-\nu(\beta)}{\beta} - \nu(\beta)
\right)+o\left(\frac{1}{n}\right).
\end{align}
These equations show that the free energy and generalization and training errors
are asymptotically given by the real log canonical threshold and  singular 
fluctuation. 
It follows that 
 \begin{align}\label{eq:WAIC}
 \EE_{X^n}[G_n(\beta)]=\EE_{X^n}\left[T_n(\beta)
 +\beta \VV_w^\beta[\log p(X_i|w)]\right]
 +o\left(\frac{1}{n}\right)
 \end{align}
 holds without explicit values of $\lambda$ and $\nu(\beta)$, 
 which shows that the difference between the generalization error 
 and the training error is asymptotically equal to the 
 posterior variance of point-wise log likelihood function. 
 Hence we can compare their theoretical values 
with the experimental ones without their explicit values. 
 
 \subsection{Real log canonical threshold} 
 \label{subsection:RLCT}
 
 In this subsection, mathematical properties and applications of the real 
 log canonical threshold are studied. The concept 
  {\it log canonical threshold} is a
 well-known birational invariant 
 in high dimensional complex 
 algebraic geometry \cite{Kollar1997}. 
 The real log canonical threshold (RLCT) is 
 the corresponding concept in real algebraic geometry 
 \cite{Saito2007,Lin2011}. 
 
Let us introduce a zeta function of $z\in\CC$,
\[
\zeta(z)=\int K(w)^z\varphi(w)dw,
\]
which was first created by Gel'fand \cite{Gelfand1964} 
and has been studied by 
many mathematicians. By the definition, 
$\zeta(z)$ is an holomorphic function in $\Re(z)>0$ which 
can be analytically continued to the unique meromorphic function
on the entire complex plane. The uniqueness of 
the analytic continuation can be proven by either 
the Hironaka resolution theorem in subsection 
\ref{subsection:mainresult} or 
the existence of {\it Bernstein-Sato polynomial}
\cite{Bernstein1971,Sato1972}. 
There exist both a differential operator $D_w$ and
a polynomial $b(z)$ such that, for arbitrary $z\in\CC$ and $w\in W$, 
\[
D_wK(w)^{z+1}=b(z)K(w)^z.
\]
Then the monic polynomial that has the lowest order and 
satisfies this equation is called 
Bernstein-Sato polynomial. By Hironaka resolution theorem, 
All poles of the zeta function is real and negative numbers. 
The largest pole and 
its order of the zeta function are
 equal to $(-\lambda)$ and $m$ respectively. 
 By using the zeta function,
 it is trivial that 
statistical learning theory of a pair 
\[
(p(x|w),\varphi(w)) 
\]
is equivalent to that of a pair 
\[
(p(x|g(u)),\varphi(g(u))|g'(u)|), 
\]
hence both $(-\lambda)$ and $m$ are birational invariants. 
 
It was also proved by Hironaka that 
there exists an algebraic algorithm by which both 
${\cal M}$ and $w=g(u)$ 
can be found by finite recursive blow-ups \cite{Hironaka1964}.
In general, such ${\cal M}$ and $w=g(u)$ are not unique. 
 If Newton diagram of $K(w)$
 is nondegenerate, they can be found by using a toric modification, which 
was applied to statistics and machine learning \cite{Yamazaki2010}. 

RLCT is determined uniquely for a given pair $(K(w),\varphi(w))$. 
There are several mathematical properties. 
\begin{enumerate}
\item 
If there exists the unique $w_0\in W_0$ such that $\nabla^2 L(w_0)$ is 
positive definite and $\varphi(w_0)>0$, then $\lambda =d/2$.
\item
If there exists $w_0\in W_0$ such that $\det(\nabla^2 L(w_0))=0$ and 
$\varphi(w_0)>0$, then $0<\lambda < d/2$.
\item
Note that Jeffreys' prior is equal to zero at singularities, and if 
Jeffreys' prior is employed in  singular  models, then $\lambda\geq d/2$. 
\item
Assume that $\lambda_j$ $(j=1,2)$ are RLCTs of $(K_j(w_j),\varphi_j(w_j))$.
Then 
\begin{itemize}
\item
RLCT of $(\sum_j K_j(w_j),\prod_j\varphi_j(w_j))$ 
is equal to $(\sum_j\lambda_j)$. 
\item
RLCT of 
$(\prod_j K_j(w_j),\prod_j\varphi_j(w_j))$ 
is equal to $(\min_j\lambda_j)$. 
\end{itemize}
\item
Assume that $\lambda_j$ $(j=1,2)$ are RLCTs of $(K_j(w),\varphi_j(w))$
and that $K_1(w)\leq c_1K_2(w)$ 
and  $\varphi_1(w)\geq c_2\varphi_2(w)$ 
for some $c_1,c_2>0$. Then $\lambda_1\leq \lambda_2$. 
\item
Let $K_1(w)=\sum_{j=1}^{J_1} f_j(w)^2$ and $K_2(w)=\sum_{j=1}^{J_2} g_j(w)^2$. If the ideal generated from $\{f_j(w)\}$ is equal to
that from $\{g_j(w)\}$, then $(K_1(w),\varphi_1(w))$ and 
$(K_2(w),\varphi_2(w))$ have the same RLCT. 

\end{enumerate}
By using these properties, 
RLCTs of important 
statistical models and learning machines were found by developing 
resolution procedures in 
neural networks \cite{Watanabe2001a,Aoyagi2012},
deep linear networks \cite{Aoyagi2024}, deep convolution ReLU network
\cite{Nagayasu2024} with and without skip connection, 
normal mixtures \cite{Yamazaki2003},
Poisson mixtures \cite{Sato2019}, multinomial mixtures \cite{WatanabeT2022}, 
general and nonnegative matrix factorization \cite{Aoyagi2005,Hayashi2017},
Boltzmann machines \cite{Yamazaki2005a}, 
hidden and general Markov models \cite{Yamazaki2005,Zwiernik2011},
and latent Dirichlet allocations \cite{Hayashi2021}.

In statistical inference, the free energy $F(1,X^n)$ $(\beta=1)$
 is often employed as 
an evaluation criterion of a pair
$(p(x|w),\varphi(w))$. 
Because numerical
calculation of the free energy requires heavy computational costs, 
several approximation methods were developed. RLCT is useful 
for such a purpose. 

If $q(x)$ is regular for $p(x|w)$, then $\lambda=d/2$ and $m=1$, 
hence
$F(1,X^n)$ is approximated by BIC \cite{Schwarz1978},
\[
{\rm BIC}=H(\hat{w},X^n)+\frac{d}{2}\log n,
\]
where $\hat{w}$ is the parameter which minimizes $H(w,X^n)$. 
In regular cases,
the difference between the free energy and BIC is a constant order 
random variable. 
For general cases when $q(x)$ may be singular for $p(x|w)$, 
the  singular BIC  was proposed \cite{Drton2017}
by using the estimated RLCT $\hat{\lambda}$, 
\[
{\rm sBIC}=H(\hat{w},X^n)+\hat{\lambda}\log n. 
\]
The difference between $F(1,X^n)$ and sBIC is smaller than any $\log n$
order random variable. 

Another method for approximation of the free energy was proposed \cite{Watanabe2013}. 
Since $F(0,X^n)=0$,
there exists $0<\beta^*<1$ such that
\begin{align}
F(1,X^n)&=\frac{\partial F}{\partial \beta}(\beta^*,X^n)
=\EE_{w}^{\beta^*}[H(w,X^n)].
\end{align}
Then we can show that
\[
\beta^*=1/\log n+o_p(1/\log n).
\]
By the definition, 
\[
{\rm WBIC}=\EE_{w}^{\beta^*}[H(w,X^n)],
\] 
the difference between 
$
F(1,X^n)$ and $WBIC$ is smaller than any 
$\log n$ order random variable. 
In the numerical calculation of WBIC, the posterior 
distribution with the inverse temperature $1/\log n$ is necessary. 
An efficient algorithm to generate such posterior distribution in mixtures 
models were proposed \cite{Watanabe2021}. 

\subsection{Singular Fluctuation}\label{subsection:SF}

In this subsection, mathematical properties and applications of
 singular  fluctuation are studied. The essential concept underlying
singular  fluctuation has been studied in many statistical contexts
\cite{Akaike1974,Amari1992,Amari1993,Murata1994}.

The original appearance of the singular fluctuation is 
model selection criteria. 
If $q(x)$ is  regular   for $p(x|w)$,  the singular   fluctuation 
is equal to 
\[
\nu(\beta)=\frac{1}{2}\mbox{tr}(IJ^{-1}),
\]
which does not depend on $\beta$,
where $I$ and $J$ are $d\times d$ matrices, 
\begin{align}
I&=\int q(x)(\nabla \log p(X|w_0))(\nabla \log p(x|w_0))dx,
\\
J&=-\nabla^2 L(w_0).
\end{align}
Moreover, if $q(x)$ is realizable by $p(x|w)$, then $I=J$, resulting that
$\nu(\beta)=d/2$ where $d$ is the dimension of $w$. 

If $q(x)$ is regular for $p(x|w)$, the limit $\beta\rightarrow\infty$ 
results in the maximum likelihood method in statistics, in which 
posterior distribution converges to the delta function on the
 maximum likelihood  estimator. Then
\[
\EE_{X^n}[G_n(\infty)]=
\EE_{X^n}[T_n(\infty)]+d/n
\]
holds, which was first found by Akaike \cite{Akaike1974}.
The $2n$ times of the right hand side of this equation is
called {\it Akaike information criterion} (AIC). If $q(x)$ is not realizable  by $p(x|w)$, 
then $d$ is replaced by $\mbox{tr}(IJ^{-1})$ which is
called by {\it Takeuchi information criterion} (TIC). 
They are 
defined as
\begin{align}
{\rm AIC}&=T_n(\infty)+d/n,
\\
{\rm TIC}&=T_n(\infty)+\mbox{tr}(IJ^{-1})/n.
\end{align}
Remark that, in practical applications, $2n$ times those values 
are used. 

For the case when $\beta=1$ and 
$q(x)$ is regular for and realizable by a model,
the $\EE_{X^n}[G(1,X^n)]$ can be estimated by deviance information
criterion (DIC) \cite{Spiegel2002}. 
The right hand side of eq.(\ref{eq:WAIC}) is called {\it WAIC} which 
can be used to estimate the generalization error  in general cases. 

The {\it leave-one-out 
cross validation} (LOOCV) \cite{Gelfand1992,Vehtari2002,Gelman2014}
is an alternative method to estimate the generalization
error. 
\[
C_n(\beta)=-\frac{1}{n}\sum_{i=1}^n \log p(X_i|X^n\setminus X_i),
\]
where
 \[
 X^n\setminus X_i=\{X_j\;;\;j=1,2,...,n\;(j\neq i)\}
 \]
 is the sample leaving $X_i$ out. Then the  generalization error
 can be estimated by $C_n(\beta)$, 
 \[
 \EE_{X^n}[C_n(\beta)]=\EE_{X^{n-1}}[G_{n-1}(\beta)]+S
 \]
Note that, if one has a numerical approximation software 
of the posterior average $\EE_w^\beta[\;\;]$ 
by using Markov chain Monte Calro method,
then $C_n(\beta)$ can be calculated by
\begin{align}\label{eq:cn}
C_n(\beta)=
-\frac{1}{n}
\sum_{i=1}^n
\log
\left(
\frac{\EE_w^\beta[p(X_i|w)^{1-\beta}]
}{
\EE_w^\beta[p(X_i|w)^{-\beta}]
}
\right).
\end{align}
The difference between LOOCV and WAIC  is smaller than any
$1/n$ order random variable. In numerical calculation of 
eq.(\ref{eq:cn}), the stability of the expectation 
$\EE_w^\beta[p(X_i|w)^{-\beta}]$
 is not ensured 
if a leverage sample point is contained in a sample 
\cite{Peruggia1997,Epifani2008}. 
A new calculation method of LOOCV was proposed by using the
approximation in MCMC distribution \cite{Vehtari2017}.
When the conditional probability $q(y|x)$ is estimated in regression problems,  
LOOCV requires that both $\{X_i\}$ and $\{(Y_i|X_i)\}$ are independent, whereas 
information criteria do that only $\{(Y_i|X_i)\}$ are independent. 
On the other hand, 
information criteria needs that the sample size $n$ is sufficiently large, 
hence LOOCV and information criteria are different statistical tools 
\cite{Watanabe2018,Watanabe2021a}. 
The concept of generalization error was extended onto 
the problems such as covariate shift, causal inference \cite{Ibayano2021},
and overparametrized cases \cite{Okunoyano2023}, and the generalized 
information criteria are now being developed. 

\subsection{Singular Learning Theory and AI Alignment}

In 1966, the statistician I. J. Good, who firstly found the 
importance of the marginal likelihood  $Z(1,X^n)$ in statistical
context, speculated as \cite{Good1966}
\begin{quotation}
``An ultra-intelligent machine is a machine that can far surpass all the intellectual activities of any man however clever. The design of machines is one of these intellectual activities; therefore, an ultra-intelligent machine could design even better machines.''
\end{quotation}
Artificial neural network research started in the 1960s.
Nowadays, 
artificial intelligence (AI) development is progressing so rapidly
 that it is difficult to predict to which direction AI takes us.
 It is said that alignment of artificial intelligence is necessary 
 for safety and welfare of human being in the near future, where 
AI alignment is sometimes defined as AI working without 
 deviating from the designer's intention.
 
 In order to consider AI alignment, we need all opinions from
 all different viewpoints, and problems to be 
 studied are beging proposed \cite{Anwar2024,Bereska2024}.
Note that singularities in learning machines make
the free energy and the generalization loss smaller 
if Bayesian statistics is employed in machine learning, 
hence degenerate property of deep
neural networks can be understood as an advantage for
constructing artificial intelligence \cite{Wei2022}. 
In AI alignment, 
 we need to study developmental landscape of in-context learning in also 
 transformers \cite{Hoogland2024}. Studying
deneracy in the loss function \cite{Lau2023} is necessary in mechanistic interpretability
\cite{Bushnaq2024}. The refined learning coefficient or RLCT is 
used for developmental interpretability \cite{Wang2025}.
 
In this subsection, we discuss AI alignment from 
the viewpoint of equivalence of 
statistical mechanics and machine learning.
There are at least three problems caused by 
singularities in artificial neural networks: 
lack of identifiability, difficulty in design of the prior distribution,
and phase transition in learning process. 
These three points originated from the hierarchical 
structure of learning machines. Remark that 
hierarchical structure is necessary for a learning machine
to have an ability of efficient universal approximation, that is to say, 
an arbitrary continuous function can be approximated by 
multi-layered neural networks with higher precision. 
Moreover it makes the generalization error
far smaller if a learning machine is over-parametrized. 
Therefore, solving the above three problems are necessary. 
I now explain these problems one by one. 

The first problem is {\it lack of identifiability}. 
A statistical model or a leaning machine
$p(x|w)$ is called {\it identifiable} if the
map $w\mapsto p(x|w)$ is one-to-one. 
Many of classical statistical models are identifiable, 
hence the role of
 the parameter is uniquely determined. Therefore, 
by checking the obtained parameter, we can examine 
the reason why a machine answers an output for a given
input. 
 However, in singular learning machines, the role of
 the parameter is not uniquely determined, hence, 
 even by checking the parameter, 
 we cannot examine whether the model's learning 
 is subject to the designer's  intention. 
 
 By introducing an equivalence relation
 \[
 w_1\sim w_2\Longleftrightarrow 
 (\forall x)\;\;p(x|w_1)=p(x|w_2),
 \]
and preparing the quotient set $W\!/\!\sim$, 
the map from an equivalence class to a probability density function
can be made one-to-one. This method is called {\it blow-down} in 
algebraic geometry. 
However, if $w_0$ is a singularity 
in the original parameter space, then its neighborhood in $W\!/\!\sim$
 has infinite dimension in general, resulting that difficulty in 
understanding the parameter cannot be solved by considering $W\!/\!\sim$.

The second problem is {\it difficulty in design of prior distribution}. 
 In the older Bayesianism in the 20th century, it was premised
that uncertainty could be captured by a statistical model and
the personal belief should be represented by a prior distribution. 
However, 
because of nonidentiability, each element of the parameter does not have 
any concrete meaning, hence the prior distribution cannot be 
determined by any {\it a priori} personal knowledge about data-generating 
process. One might think that Jeffreys' prior may be chosen
as an objective one, however, it makes the generalization error
far larger than others \cite{Watanabe2000}, which destroys the advantage of hierarchical structure. 

We need modern Bayesian statistics,  in which a statistical
 model or a learning machine 
 is only a candidate and that a prior distribution is not
 any belief of a person 
 but a part of a statistical model. 
 In the practical applications, ridge or lasso type prior distribution
  with hyperparameter 
 is sometimes employed and the hyperparameter is optimized so that
 the generalization error or the free energy is made smaller
 by several criteria including cross validation. 
 
The third problem is the {\it phase transition phenomenon in 
learning process} \cite{Watanabe2001}. In classical statistical models, 
the optimal parameter $w_0$ is unique and the posterior
distribution can be approximated as a normal distribution 
which converges to the delta function $\delta(w-w_0)$ as 
the sample size increases and 
the posterior distribution may be understood as a gradual increase of 
confidence. 

On the other hand, in deep learning, 
 even if the sample size
 is huge, it is far smaller than 
 the infinity for the deep neural network.
Let $q(x)$ be unrealizable by $p(x|w)$. 
By using division of unity, there is a finite set of 
nonngegative $C_0^\infty$ class functions $\{\varphi_k(w)\}$ such that
\[
\varphi(w)=\sum_{k=1}^K\varphi_k(w)
\]
Then it follows that
\[
Z_n(\beta)=\sum_{k=1}^K Z_{nk}(\beta),
\]
where 
\[
Z_{nk}(\beta)=\int \exp(-\beta H(w,X^n))\varphi_k(w)dw.
\]
Let $w_k$ be the minimum point of $L(w)$ in the
support of $\varphi_k(w)$ and $\lambda_k$ be the 
real log canonical threshold for $(q(x),p(x|w),\varphi_k(w))$.
The local free energy is given by
\[
F_{nk}(\beta)=
nL(w_k)+\frac{\lambda_k}{\beta}\log n +O(\log \log n) .
\]
Then by using an inequality 
\[
\min_{k} F_{nk}(\beta) -\log K
\leq 
-\log\left( \sum_{k=1}^K\exp(-F_{nk}(\beta))\right)
\leq 
\min_{k} F_{nk}(\beta), 
\]
it follows that 
\[
F_n(\beta)=\min_k\left\{
nL(w_k)+\frac{\lambda_k}{\beta}\log n\right\}+O(\log\log n).
 \]
 Therefore the free energy is given by the minimum of the
 local free energy. In statistical learning theory, $L(w_k)$ and
 $\lambda_k\log n$ are called {\it bias} and {\it variance}, respectively. 
Note that $\exp(-F_{nk})$ is in proportion to the 
posterior probability of the local coordinate. 
Therefore the optimal local parameter set for minimum free energy
 is automatically
chosen by the posterior probability. However, it is not equal to the
optimal parameter set for the minimum generalization error. 
The local parameter set chosen by the posterior distribution
jumps from a neighborhood to another one as sample size increases. 
 Such jumping process makes it 
 difficult to diagnose whether 
 the learning process is out of 
 order or not. This is a phase transition phenomenon with respect to
 the increase of the sample size. 

Other
phase transition phenomena can be also observed in different situations. 
In mixture models, a Dirichlet distribution is often chosen for a prior 
distribution of the mixture ratio. Then the posterior distribution has a 
phase transition according to the hyperparameter of Dirichlet distribution
\cite{Watanabe2018,WatanabeT2022}. In the different phases, RCLTs are
different and the supports of asymptotic posterior distributions are different,
which means that the local parameter sets for the different hyperparameters
may be distant from each other \cite{Watanabe2018}.

\section{Conclusion}

In this paper, 
for theoretical physicists who are interested in artificial intelligence,
we introduced 
mathematical equivalence between statistical mechanics of
random Hamiltonian and machine learning theory. 
Professor Huzihiro Araki indicated that algebraic concepts such as
operator algebras play important roles in statistical mechanics and
quantum field theory. In this paper, we proposed that, also in machine
learning theory, algebraic research is useful in both fundamental 
and practical fields. For example, such researches are becoming more
important in 
developing statistics and artificial intelligence. 

\section*{Acknowledgment}

The author would like to express his sincere gratitude to Professor Huzihiro
Araki who is a pioneering founder of algebraic research  in both 
statistical mechanics and quantum field theory.

\end{document}